\documentstyle[preprint,aps]{revtex}
\tightenlines

\begin{document}
\draft
\title{Two-particle model with noncommuting operators of coordinates and momenta}
\author{M. V. Kuzmenko}
\address{Bogolyubov Institute for Theoretical Physics of the NAS of Ukraine, \\
Metrolohichna Str., 14b, Kyiv-143, 03143 Ukraine}
\date{\today}
\maketitle

\begin{abstract}
A nonrelativistic equation for the system of two interacting particles
within the framework of a model with noncommuting operators of coordinates
and momenta of different particles is proposed, and a self-consistent system
of equations for the wave function of every quantum state is deduced.
Solutions for the lowest states of a hydrogenlike atom are found, and the
comparison with analogous solutions of the Klein-Gordon equation for the
relativistic spinless problem is performed. In the case where the size of a
two-particle system and the Compton wavelengths of particles forming it are
of the same order, the essential differences with solutions of the
Schr\"{o}dinger nonrelativistic equation are revealed.
\end{abstract}

\pacs{03.65.Ge, 03.65.Bz, 31.10.+z}

\section{Introduction}

For a system of two interacting particles, the well-known Schr\"{o}dinger
equation \cite{R1} can be derived through the formal substitution of the
quantities $E$, ${\bf r}_{1}$, ${\bf r}_{2}$, ${\bf p}_{1}$, and ${\bf p}%
_{2} $ by the relevant operators \cite{R2} on both sides of the relation $%
E=H({\bf r}_{1}{\bf ,r}_{2}{\bf ,p}_{1},{\bf p}_{2})$, where $H$ is the
classical Hamilton function of a system of two interacting particles. Here, $%
{\bf r}_{1}$ and ${\bf r}_{2}$ are Cartesian coordinates of two particles
with momenta ${\bf p}_{1}$ and ${\bf p}_{2}$, respectively.

The operators ${\bf \hat{r}}_{1}$, ${\bf \hat{r}}_{2}$, ${\bf \hat{p}}_{1}$,
and ${\bf \hat{p}}_{2}$ are such that they satisfy the following commutation
relations:
\begin{equation}
\left[ \hat{x}_{k},\,\hat{p}_{kx}\right] =\left[ \hat{y}_{k},\,\hat{p}_{ky}%
\right] =\left[ \hat{z}_{k},\,\hat{p}_{kz}\right] =i\hbar \,\text{\ }%
(k=1,\,2)\,.  \label{E1}
\end{equation}
Here, $\hbar =h/2\pi $, where $h$ is the universal constant introduced by
Planck. All other possible commutation relations equal zero, including
\begin{equation}
\left[ \hat{x}_{k},\,\hat{p}_{lx}\right] =\left[ \hat{y}_{k},\,\hat{p}_{ly}%
\right] =\left[ \hat{z}_{k},\,\hat{p}_{lz}\right] =0\,\text{\ }(k\neq l)\,.
\label{E2}
\end{equation}

Equalities~(\ref{E2}) are based on the assumption that measurements of
coordinates and momenta of different particles do not disturb one another in
principle even in the presence of some forces between particles~ \cite{R3}.
That is, one supposes that the change in the force action of a particle on
another one, caused by a measurement of the coordinate of the first,
propagates with finite velocity.

Thus, to derive the Schr\"{o}dinger nonrelativistic equation for a
two-particle system, one uses, on the one hand, the Hamilton classical
nonrelativistic function and, on the other hand, the implicit assumption
about finiteness of the interaction propagation velocity.

In the fully nonrelativistic quantum theory, we must consider the
interaction propagation velocity as infinitely large, which forces us to
drop the requirement for the commutation relations~(\ref{E2}) to hold.
Having accepted this viewpoint, we will consider that, under a measurement
of the coordinate of the first particle, there occurs the uncontrolled
transfer of momentum not only to this particle but to the whole system since
the particles are connected through the interaction potential, whose
propagation velocity is infinitely large.

Therefore, it is natural to require that the commutator of the coordinate
operator of any particle and the of operator total momentum of the system $%
{\bf \hat{P}}_{c}={\bf \hat{p}}_{1}+{\bf \hat{p}}_{2}$ be equal to $i\hbar $%
:
\begin{equation}
\left[ \hat{x}_{k},\,\hat{P}_{cx}\right] =\left[ \hat{y}_{k},\,\hat{P}_{cy}%
\right] =\left[ \hat{z}_{k},\,\hat{P}_{cz}\right] =i\hbar \,\text{\ }(k=1,2).
\label{E3}
\end{equation}
Note that relations~(\ref{E3}) hold true also for a Schr\"{o}dinger
nonrelativistic equation. Namely, they allow one to construct the operator
of coordinates of the center of mass of the system, whose commutator with
the operator of total momentum equals $i\hbar $. On the contrary, the
fulfillment of relations~(\ref{E1}) is not obligatory for a system of
interacting particles, and we intend to reject this requirement.

The first attempt to construct the nonrelativistic equation for a system of
interacting particles in the framework of the model with noncommuting
operators of coordinates and momenta of different particles was undertaken
in \cite{R4,R5}. However, the theory proposed there is phenomenological,
i.e., it contains a parameter whose determination is ambiguous. Moreover,
the proposed self-consistent system of integro-differential equations was
faced with difficulties of the probabilistic interpretation of a wave
function. The present work is devoted to the elimination of the drawbacks
mentioned above.

\section{Fully nonrelativistic statement of the quantum two-body problem}

The probabilistic interpretation of the square of the modulus of a wave
function is possible only under the assumption that measurements of
coordinates or momenta of various particles do not principally disturb one
another even if there exists some interaction between particles~\cite{R3}.
This means that the operators of coordinates and momenta of two particles
commute with each other.

As noted above, the operators of coordinates and momenta of different
particles do not commute with one another in the general case. Let
\begin{equation}
\left[ \hat{x}_{1},\,\hat{p}_{2x}\right] =i\hbar \hat{f}_{1}\,,  \label{E4}
\end{equation}
where $\hat{f}_{1}$ is a dimensionless Hermitian operator. Then it follows
from Eq.~(\ref{E3}) that
\begin{equation}
\left[ \hat{x}_{1},\,\hat{p}_{1x}\right] =i\hbar (1-\hat{f}_{1})\,.
\label{E5}
\end{equation}
By analogy, if
\begin{equation}
\left[ \hat{x}_{2},\,\hat{p}_{1x}\right] =i\hbar \hat{f}_{2}\,,  \label{E6}
\end{equation}
then
\begin{equation}
\left[ \hat{x}_{2},\,\hat{p}_{2x}\right] =i\hbar (1-\hat{f}_{2})\,.
\label{E7}
\end{equation}
The dimensionless Hermitian operators $\hat{f}_{1}$ and $\hat{f}_{2}$ depend
generally on the interaction force between particles ${\bf F}_{12}$ and on
the masses of interacting particles $m_{1}$ and $m_{2}$. The operators $\hat{%
f}_{1}$ and $\hat{f}_{2}$ cannot depend on a direction of the vector ${\bf F}%
_{12}$, since the commutation relations for the $x$, $y$, and $z$ components
should be identical analogously to~(\ref{E4})-(\ref{E7}), since there are no
separated directions in the system, and the independent variables in the
Cartesian coordinate system are fully equivalent. For this reason, the
operators $\hat{f}_{1}$ and $\hat{f}_{2}$ are only functions of the absolute
value of a force, i.e., of $\left| {\bf F}_{12}\right| $:
\begin{equation}
\hat{f}_{1}\equiv \hat{f}_{1}(m_{1},m_{2},\left| {\bf F}_{12}\right| )\,,%
\text{\quad }\hat{f}_{2}\equiv \hat{f}_{2}(m_{1},m_{2},\left| {\bf F}%
_{12}\right| )\,.  \label{E8}
\end{equation}
Let us make permutation of $m_{1}$ and $m_{2}$. Then
\begin{equation}
\left[ \hat{x}_{1},\,\hat{p}_{2x}\right] =i\hbar \hat{f}_{1}(m_{2},m_{1},%
\left| {\bf F}_{12}\right| )\,,\text{\quad }\left[ \hat{x}_{2},\,\hat{p}_{1x}%
\right] =i\hbar \hat{f}_{2}(m_{2},m_{1},\left| {\bf F}_{12}\right| )\,.
\label{E9}
\end{equation}
Compare~(\ref{E9}) with~(\ref{E4}), (\ref{E6}). Considering that the
physical situation has not changed, we get
\begin{equation}
\hat{f}_{1}(m_{2},m_{1},\left| {\bf F}_{12}\right| )=\hat{f}%
_{2}(m_{1},m_{2},\left| {\bf F}_{12}\right| )\,.  \label{E10}
\end{equation}

Thus, we have one unknown operator $\hat{f}_{1}(m_{1},m_{2},\left| {\bf F}%
_{12}\right| )$. For $m_{2}\rightarrow 0$, $\hat{f}_{1}$ must tend to zero
since, in the absence of the second particle, the whole momentum transferred
under the measurement of the coordinate $x_{1}$ falls to the first one. If $%
\left| {\bf F}_{12}\right| \rightarrow 0$, then $\hat{f}_{1}\rightarrow 0$,
i.e., without any interaction forces between particles, the operators of
coordinates and momenta of different particles commute among themselves. The
situation $\left| {\bf F}_{12}\right| \rightarrow \infty $ corresponds to
the case where we have one particle of mass $M$ formed by two strongly bound
particles with masses $m_{1}$ and $m_{2}$. Therefore, the momentum, received
under a measurement of some coordinate, is distributed proportionally to
masses of particles. This enables us to write down $\hat{f}_{1}$ as $\hat{f}%
_{1}=m_{2}/M$. Here, $M=m_{1}+m_{2}$ is the system mass.

Therefore, without loss of generality, we can present the operator $\hat{f}%
_{1}$ as
\begin{equation}
\hat{f}_{1}=\frac{m_{2}}{M}\hat{\varepsilon}(\left| {\bf F}_{12}\right|
,m_{1},m_{2})\,,  \label{E11}
\end{equation}
where $\hat{\varepsilon}$ is a new operator, which is assumed to be
symmetric with respect to the masses of particles $m_{1}$ and $m_{2}$. In
what follows, we will omit its explicit dependence on masses to shorten
formulas, namely, $\hat{\varepsilon}(\left| {\bf F}_{12}\right|
,m_{1},m_{2})\equiv \hat{\varepsilon}(\left| {\bf F}_{12}\right| )$. For $%
\left| {\bf F}_{12}\right| \rightarrow 0$, $\hat{\varepsilon}\rightarrow 0$,
and $\hat{\varepsilon}\rightarrow 1$ for $\left| {\bf F}_{12}\right|
\rightarrow \infty $.

For the noncommuting operators $\hat{x}_{1}$ and $\hat{p}_{2x}$, the
uncertainty relation has the form~
\begin{equation}
\Delta x_{1}\,\Delta p_{2x}\geq \frac{\hbar }{2}\frac{m_{2}}{M}\left| \frac{%
\left\langle \Psi \left| \hat{\varepsilon}(\left| {\bf F}_{12}\right|
)\right| \Psi \right\rangle }{\left\langle \Psi \left| {}\right. \Psi
\right\rangle }\right| \,,  \label{E12}
\end{equation}
where $\left\langle \Psi \left| \hat{\varepsilon}(\left| {\bf F}_{12}\right|
)\right| \Psi \right\rangle \left( \left\langle \Psi \left| {}\right. \Psi
\right\rangle \right) ^{-1}\equiv \varepsilon $ is a quantum-mechanical
average of the operator $\hat{\varepsilon}$ in the state $\Psi ({\bf r}_{1}%
{\bf ,r}_{2}{\bf ,}t).$

Assume now that the two-particle system has only one quantum state $\Psi
_{1}({\bf r}_{1}{\bf ,r}_{2}{\bf ,}t)$. If we substitute the operator $\hat{%
\varepsilon}$ in the formula for $\hat{f}_{1}$ by its quantum-mechanical
average $\left\langle \Psi _{1}\left| \hat{\varepsilon}(\left| {\bf F}%
_{12}\right| )\right| \Psi _{1}\right\rangle \left( \left\langle \Psi
_{1}\left| {}\right. \Psi _{1}\right\rangle \right) ^{-1}\equiv \varepsilon
_{1}$, the uncertainty relation (\ref{E12}) does not change. This allows us
to construct a nonrelativistic equation for the two-particle system since
the operator $\hat{f}_{1}$ is constant now.

We present the commutation relations for all operators of coordinates and
momenta in the two-body problem (for the y and z components, the commutators
are the same) as
\begin{equation}
\left[ \hat{x}_{1},\,\hat{p}_{1x}\right] =i\hbar \left( 1-\frac{m_{2}}{M}%
\varepsilon _{1}\right) \,,  \label{E13}
\end{equation}
\begin{equation}
\left[ \hat{x}_{2},\,\hat{p}_{2x}\right] =i\hbar \left( 1-\frac{m_{1}}{M}%
\varepsilon _{1}\right) \,,  \label{E14}
\end{equation}
\begin{equation}
\left[ \hat{x}_{1},\,\hat{p}_{2x}\right] =i\hbar \frac{m_{2}}{M}\varepsilon
_{1}\,,  \label{E15}
\end{equation}
\begin{equation}
\left[ \hat{x}_{2},\,\hat{p}_{1x}\right] =i\hbar \frac{m_{1}}{M}\varepsilon
_{1}\,,  \label{E16}
\end{equation}
\begin{equation}
\left[ \hat{x}_{1},\,\hat{x}_{2}\right] =0\,,  \label{E17}
\end{equation}
\begin{equation}
\left[ \hat{p}_{1x},\,\hat{p}_{2x}\right] =0\,.  \label{E18}
\end{equation}

Now we can construct one of the possible representations for the operators
of coordinates and momenta of the two-particle system as
\begin{equation}
{\bf \hat{r}}_{1}={\bf r}_{1}{\bf \,\,,}  \label{E19}
\end{equation}
\begin{equation}
{\bf \hat{r}}_{2}={\bf r}_{2}{\bf \,\,,}  \label{E20}
\end{equation}
\begin{equation}
{\bf \hat{p}}_{1}=-i\hbar \left( 1-\frac{m_{2}}{M}\varepsilon _{1}\right)
{\bf \nabla }_{{\bf 1}}-i\hbar \frac{m_{1}}{M}\varepsilon _{1}{\bf \nabla }_{%
{\bf 2}}\,,  \label{E21}
\end{equation}
\begin{equation}
{\bf \hat{p}}_{2}=-i\hbar \frac{m_{2}}{M}\varepsilon _{1}{\bf \nabla }_{{\bf %
1}}-i\hbar \left( 1-\frac{m_{1}}{M}\varepsilon _{1}\right) {\bf \nabla }_{%
{\bf 2}}\,.  \label{E22}
\end{equation}
Here, as independent variables, we use the coordinates of the particles $%
{\bf r}_{1}{\bf \,}$ and ${\bf r}_{2}{\bf \,}$, since the corresponding
operators commute with one another.

By changing operators (\ref{E19})-(\ref{E22}) in the Hamilton function, we
get the self-consistent system of integro-differential equations for the
nonrelativistic two-body problem:
\begin{equation}
i\hbar \frac{\partial }{\partial t}\Psi _{1}({\bf r}_{1}{\bf ,r}_{2}{\bf ,}%
t)=\hat{H}\text{ }\Psi _{1}({\bf r}_{1}{\bf ,r}_{2}{\bf ,}t),  \label{E23}
\end{equation}
\begin{equation}
\varepsilon _{1}=\frac{\left\langle \Psi _{1}\left| \hat{\varepsilon}(\left|
{\bf F}_{12}\right| )\right| \Psi _{1}\right\rangle }{\left\langle \Psi
_{1}\left| {}\right. \Psi _{1}\right\rangle },  \label{E24}
\end{equation}
where
\begin{equation}
\hat{H}=\hat{T}+\hat{V}(\left| {\bf r}_{2}-{\bf r}_{1}\right| ),  \label{E25}
\end{equation}
\begin{eqnarray}
\hat{T} &=&-\frac{\hbar ^{2}}{2m_{1}}\left[ \left( 1-\frac{m_{2}}{M}%
\varepsilon _{1}\right) ^{2}+\frac{m_{1}m_{2}}{M^{2}}\varepsilon _{1}^{2}%
\right] \Delta _{1}-\frac{\hbar ^{2}}{2m_{2}}\left[ \left( 1-\frac{m_{1}}{M}%
\varepsilon _{1}\right) ^{2}+\frac{m_{1}m_{2}}{M^{2}}\varepsilon _{1}^{2}%
\right] \Delta _{2}  \nonumber \\
&&-\frac{\hbar ^{2}}{2M}\left[ 4\varepsilon _{1}-2\varepsilon _{1}^{2}\right]
({\bf \nabla }_{{\bf 1}}\cdot {\bf \nabla }_{{\bf 2}}).  \label{E26}
\end{eqnarray}
We recall that we have received the self-consistent system of equations (\ref
{E23})-(\ref{E26}) for the two-particle system by assuming the existence of
only one quantum state $\Psi _{1}({\bf r}_{1}{\bf ,r}_{2}{\bf ,}t)$.

Under the standard change of variables
\begin{equation}
{\bf r}={\bf r}_{2}-{\bf r}_{1},  \label{E27}
\end{equation}
\begin{equation}
{\bf R}=\frac{m_{1}{\bf r}_{1}+m_{2}{\bf r}_{2}}{m_{1}+m_{2}},  \label{E28}
\end{equation}
where ${\bf r}$ and ${\bf R}$ are the relative distance between particles
and the coordinates of the center of mass of the system, the system of
equations (\ref{E23})-(\ref{E26}) becomes simpler:
\begin{equation}
{i\hbar \frac{\partial }{\partial t}}\Psi _{1}({\bf r,R,}t){=\left[ -\frac{%
\hbar ^{2}}{2M}\Delta _{{\bf R}}-\frac{\hbar ^{2}(1-\varepsilon _{1})^{2}}{%
2\mu }\Delta _{{\bf r}}+V\left( r\right) \right] }\Psi {_{1}(}{\bf r}{\bf ,}%
{\bf R}{\bf ,}t{),}  \label{E29}
\end{equation}
\begin{equation}
\varepsilon _{1}=\frac{\left\langle \Psi _{1}\left| \hat{\varepsilon}(\left|
{\bf F}_{12}(r)\right| )\right| \Psi _{1}\right\rangle }{\left\langle \Psi
_{1}\left| {}\right. \Psi _{1}\right\rangle }\,.  \label{E30}
\end{equation}
Here, $\mu $ is the reduced mass, $\mu =m_{1\,}m_{2}/(m_{1}+m_{2})$. The
operator of total momentum reads
\begin{equation}
{\bf \hat{P}}_{c}={\bf \hat{p}}_{1}+{\bf \hat{p}}_{2}=-i\hbar {\bf \nabla }_{%
{\bf R}}\,.  \label{E31}
\end{equation}

Let the Hamiltonian $H$ do not explicitly depend on time. Then we obtain, by
the substitution
\begin{equation}
\Psi _{1}=\psi _{1}\exp \left( -i\frac{Et}{{\hbar }}\right) \,,  \label{E32}
\end{equation}
where $\psi _{1}$ depends on coordinates in the configuration space but not
on time, the self-consistent system of integro-differential equations for a
stationary state of the two-particle system as
\begin{equation}
{\left[ -\frac{\hbar ^{2}}{2M}\Delta _{{\bf R}}-\frac{\hbar
^{2}(1-\varepsilon _{1})^{2}}{2\mu }\Delta _{{\bf r}}+V(r)\right] }\psi {%
_{1}(}{\bf r}{\bf ,}{\bf R}{)=}E\psi {_{1}(}{\bf r}{\bf ,}{\bf R}{),}
\label{E33}
\end{equation}
\begin{equation}
\varepsilon _{1}=\frac{\left\langle \psi _{1}\left| \hat{\varepsilon}(\left|
{\bf F}_{12}(r)\right| )\right| \psi _{1}\right\rangle }{\left\langle \psi
_{1}\left| {}\right. \psi _{1}\right\rangle }\,.  \label{E34}
\end{equation}

By the substitution $\psi _{1}({\bf r,R)=}\Phi {\bf (R)\,}\varphi _{1}{\bf %
(r)}$, we can separate the motion of the center of mass of the system as a
whole. As a result, we arrive at the following self-consistent system of
equations:
\begin{equation}
{\left[ -\frac{\hbar ^{2}(1-\varepsilon _{1})^{2}}{2\mu }\Delta _{{\bf r}%
}+V\left( r\right) \right] }\varphi {_{1}{\bf (r)}=}E{_{1}}\varphi {_{1}{\bf %
(r)},}  \label{E35}
\end{equation}
\begin{equation}
\varepsilon _{1}=\frac{\left\langle {{\bf \varphi }_{1}}\left| \hat{%
\varepsilon}(\left| {\bf F}_{12}(r)\right| )\right| {{\bf \varphi }_{1}}%
\right\rangle }{\left\langle {{\bf \varphi }_{1}}\left| {}\right. {{\bf %
\varphi }_{1}}\right\rangle }\,.  \label{E36}
\end{equation}

As in the Schr\"{o}dinger nonrelativistic theory, a wave function $\varphi
_{1}({\bf r)}$ should be continuous together with its partial derivatives of
the first order in the whole space and, in addition, be a bounded
single-valued function of its arguments.

As in the Schr\"{o}dinger theory, for particles interacting by means of a
spherically symmetric potential, which depends only on the distance between
particles, the wave function $\varphi _{1}({\bf r)}$ can be represented in
the following form:
\begin{equation}
\varphi _{1}({\bf r)=}\frac{1}{r}\chi _{1l}(r)Y_{lm}\left( \frac{{\bf r}}{r}%
\right) \,,  \label{E37}
\end{equation}
where $Y_{lm}\left( {\bf r/}r\right) $ are orthonormalized spherical
functions. Then the function $\chi _{1l}(r)$ satisfies the following system
of equations:
\begin{equation}
\left[ -\frac{\hbar ^{2}(1-\varepsilon _{1l})^{2}}{2\mu }\left( \frac{d^{2}}{%
dr^{2}}-\frac{l(l+1)}{r^{2}}\right) +V(r)\right] \chi _{1l}(r)=E_{1l}\chi
_{1l}(r)\,{\bf ,}  \label{E38}
\end{equation}
\begin{equation}
\varepsilon _{1l}=\frac{\left\langle \chi _{1l}\left| \hat{\varepsilon}%
(\left| {\bf F}_{12}(r)\right| )\right| \chi _{1l}\right\rangle }{%
\left\langle \chi _{1l}\left| {}\right. \chi _{1l}\right\rangle }\,.
\label{E39}
\end{equation}
We emphasize that, for a given value of $l$, the corresponding value of $%
\varepsilon _{1l}$ can exist.

Now we consider the laws of conservation which exist for the proposed fully
nonrelativistic scheme of the system of two interacting particles. Since no
external fields act on the system, its Hamiltonian must be invariant
relative to a parallel translation of the coordinate system in space and
time as well as relative to a rotation of the coordinate axes. In addition,
the motion equations do not change under a uniform and rectilinear motion of
the system (the Galilei invariance).

The Hamiltonian of the isolated system (\ref{E25}) does not depend on time
explicitly, and therefore the energy of the system is an integral of motion.

The operator of total momentum of the system ${\bf \hat{P}}_{c}={\bf \hat{p}}%
_{1}+{\bf \hat{p}}_{2}=-i\hbar {\bf \nabla }_{1}-i\hbar {\bf \nabla }_{2}$
is connected with the operator of infinitesimal translation, which
transforms the function $\Psi ({\bf r}_{1}{\bf ,r}_{2})$ into the function $%
\Psi ({\bf r}_{1}+\delta {\bf r,r}_{2}+\delta {\bf r})$, by the relation
\begin{equation}
1+\delta {\bf r\cdot }\sum_{k=1}^{2}{\bf \nabla }_{k}=1+\frac{i}{\hbar }%
\delta {\bf r\cdot \hat{P}}_{c}  \label{E40}
\end{equation}
and commutes with Hamiltonian (\ref{E25}):
\begin{equation}
\left[ \hat{H},\,{\bf \hat{P}}_{c}\right] =0,  \label{E41}
\end{equation}
where $\delta {\bf r}$ is the vector of infinitesimal parallel displacement
of all radius-vectors by the same quantity: ${\bf r}_{k}{\bf \rightarrow r}%
_{k}+\delta {\bf r}$. Thus, three components of the total momentum are
integrals of motion, and the total momentum of the two-particle system is
conserved.

By virtue of the isotropy of space, the Hamiltonian of a closed system must
not vary under a rotation of the whole system by an arbitrary angle around
an arbitrary axis. It is sufficient to require the fulfilment of this
condition for any infinitesimal rotation, whose vector $\delta {\bf \Omega }
$ is equal to the rotation angle $\delta \Omega $ in modulus and is
directed along the rotation axis. The operator of infinitesimal rotation,
which transforms the function $\Psi ({\bf r}_{1}{\bf ,r}_{2})$ into $\Psi (%
{\bf r}_{1}+[\delta {\bf \Omega }\times {\bf r}_{1}]{\bf ,r}_{2}+[\delta
{\bf \Omega }\times {\bf r}_{2}])$, is connected with the operator of total
angular momentum of the system through the relation
\begin{equation}
1+\delta {\bf \Omega \cdot }\sum_{k=1}^{2}[{\bf r}_{k}\times {\bf \nabla }%
_{k}]=1+\frac{i}{\hbar }\delta {\bf \Omega \cdot }{\bf \hat{L}}  \label{E42}
\end{equation}
and commutes with Hamiltonian (\ref{E25}) of the system. Thus, the total
angular momentum ${\bf \hat{L}=-}i\hbar \sum_{k=1}^{2}[{\bf r}_{k}\times
{\bf \nabla }_{k}]$ of the two-particle system is conserved.

It is worth noting the following. If we write the operators ${\bf \hat{l}}%
_{1}{\bf =}[{\bf \hat{r}}_{1}\times {\bf \hat{p}}_{1}]$ and ${\bf \hat{l}}%
_{2}{\bf =}[{\bf \hat{r}}_{2}\times {\bf \hat{p}}_{2}]$ in a formal way for
each particle, it is easy to show that they and their sum are not angular
momenta because they do not satisfy the standard commutation relations
peculiar to angular momentum:
\begin{equation}
\left[ L_{x},L_{y}\right] =i\hbar L_{z},\text{ \ \ }\left[ L_{y},L_{z}\right]
=i\hbar L_{x},\text{ \ \ \ }\left[ L_{z},L_{x}\right] =i\hbar L_{y}\,\text{\
}\,.  \label{E43}
\end{equation}
However, from the operators ${\bf \hat{r}}_{k}$ and ${\bf \hat{p}}_{n}$, one
can construct the operator possessing the mentioned properties of total
angular momentum:
\begin{equation}
{\bf \hat{L}=}\sum_{k,n}C_{kn}[{\bf \hat{r}}_{k}\times {\bf \hat{p}}_{n}]=%
{\bf -}i\hbar \sum_{k=1}^{2}[{\bf r}_{k}\times {\bf \nabla }_{k}].
\label{E44}
\end{equation}
For the two-particle system, the coefficients $C_{kn}$ are as follows:
\begin{equation}
C_{11}=\left( 1-\frac{m_{1}}{M}\varepsilon _{1}\right) (1-\varepsilon
_{1})^{-1},  \label{E45}
\end{equation}
\begin{equation}
C_{12}=-\frac{m_{1}}{M}\varepsilon _{1}(1-\varepsilon _{1})^{-1},
\label{E46}
\end{equation}
\begin{equation}
C_{21}=-\frac{m_{2}}{M}\varepsilon _{1}(1-\varepsilon _{1})^{-1},
\label{E47}
\end{equation}
\begin{equation}
C_{22}=\left( 1-\frac{m_{2}}{M}\varepsilon _{1}\right) (1-\varepsilon
_{1})^{-1}.  \label{E48}
\end{equation}

It is necessary to emphasize that the parameter of noncommutativity of the
operators of coordinates and momenta of different particles $\varepsilon
_{1} $ is the quantum-mechanical average of an operator which depends on the
modulus of the interaction force between two particles (i.e., on the
distance between two particles). Therefore, this operator is independent of
motion of the center of mass of the system since the system of equations (%
\ref{E23})-(\ref{E26}) admits the separation of motion of the center of mass
of the two-particle system as a whole.

Upon the motion of two reference systems relative to one another with
constant velocity ${\bf v}$, the operators ${\bf \hat{r}}_{k}$ and ${\bf
\hat{p}}_{k}$ are transformed into ${\bf \hat{r}}_{k}-{\bf v}t$ and ${\bf
\hat{p}}_{k}-m_{k}{\bf v}$. Such a Galilei transformation of the system of
particles is described by the operator \cite{R2}
\begin{equation}
\hat{G}({\bf v,}t)=\exp \left[ i{\bf v\cdot }\left( M{\bf \hat{R}-\hat{P}}%
_{c}t\right) /\hbar \right] ,  \label{E49}
\end{equation}
where $M$, ${\bf \hat{R}}$, and ${\bf \hat{P}}_{c}$ are the mass, operator
of coordinate, and operator of momentum of the center of mass of the
two-particle system. It is easy to show that the condition for the Galilei
invariance of the equation of motion (\ref{E23})
\begin{equation}
\hat{G}^{-1}({\bf v,}t)\left[ i\hbar \frac{\partial }{\partial t}-\hat{H}%
\right] \hat{G}({\bf v,}t)=\left[ i\hbar \frac{\partial }{\partial t}-\hat{H}%
\right]  \label{E50}
\end{equation}
is valid for Hamiltonian (\ref{E25}).

The quantity $\varepsilon _{1}m_{2}/M$ is the average share of the momentum
transferred to the second particle under the measurement of the coordinate
of the first. The value of this momentum can be estimated as $\left| {\bf F}%
_{12}\right| \Delta t$, where $\Delta t$ is the duration of measurement of
the coordinate of the first particle. Therefore, the operator $\hat{%
\varepsilon}$ can be presented in the form

\begin{equation}
\hat{\varepsilon}(\left| {\bf F}_{12}\right| )=\frac{\left| {\bf F}%
_{12}\right| \Delta t}{\left| {\bf F}_{12}\right| \Delta t+P_{0}}=\frac{%
\left| {\bf F}_{12}\right| }{\left| {\bf F}_{12}\right| +F_{0}},  \label{E51}
\end{equation}
where $P_{0}$ or $F_{0}$ is some constant of the relevant dimensional. For
the two-particle system, we can construct the quantity $F_{0}$ from two
constants $\mu $ and $\hbar $ appearing in the problem under study as
\begin{equation}
F_{0}=\frac{\left( \mu c^{2}\right) ^{2}}{\hbar c},  \label{E52}
\end{equation}
i.e., this is the ratio of the energy of rest of the particle with reduced
mass $\mu $ to its Compton wavelength. Therefore, the parameter of
noncommutativity of the operators of coordinates and momenta of different
particles can be presented in the form:
\begin{equation}
\varepsilon _{1l}=\,\left( \int_{0}^{\infty }\chi _{1l}^{2}(r)\frac{\left|
{\bf F}_{12}(r)\right| }{\left| {\bf F}_{12}(r)\right| +{\displaystyle{\frac{%
\left( \mu c^{2}\right) ^{2}}{\hbar c}}}}dr\,\right) \left( \int_{0}^{\infty
}\chi _{1l}^{2}(r)dr\,\right) ^{-1}.  \label{E53}
\end{equation}

The proposed method of determination of the parameter of noncommutativity of
the operators of coordinates and momenta of different particles contains no
parameters and therefore substantially differs from the phenomenological
method, proposed in \cite{R4}. In the latter method, the representation of
the operator $\hat{\varepsilon}=1-\exp \left( -\Omega _{0}F_{12}^{2}\left(
\left| {\bf r}_{2}-{\bf r}_{1}\right| \right) \right) $ includes the
parameter $\Omega _{0}$, whose determination is ambiguous.

Now we consider the situation where a two-particle system has $N$ quantum
states, each of them is characterized by the wave function $\Psi _{1},\ldots
,\Psi _{N}$. Let the energy of every stationary state be denoted by $%
E_{1},\ldots ,E_{N}$, respectively. If we take the parameter of
noncommutativity $\varepsilon $ to be the same for each state and equal to,
for example, the maximum value of
\begin{equation}
\varepsilon =\max \left\{ \frac{\left\langle \Psi _{1}\left| \hat{\varepsilon%
}(\left| {\bf F}_{12}\right| )\right| \Psi _{1}\right\rangle }{\left\langle
\Psi _{1}\left| {}\right. \Psi _{1}\right\rangle },\ldots ,\frac{%
\left\langle \Psi _{N}\left| \hat{\varepsilon}(\left| {\bf F}_{12}\right|
)\right| \Psi _{N}\right\rangle }{\left\langle \Psi _{N}\left| {}\right.
\Psi _{N}\right\rangle }\right\} ,  \label{E54}
\end{equation}
then operators (\ref{E21})-(\ref{E22}) will be linear. In this case, the
principle of superposition is valid for states. The probabilistic
interpretation of wave functions is also conserved.

However, such a choice of the parameter of noncommutativity $\varepsilon $
is not physically correct because there exist different average values of
the modulus of the interaction force between particles in different quantum
states, and the parameter of noncommutativity should be generally different
in each quantum state. In this case, we have the own self-consistent system
of equations of the type of (\ref{E29})-(\ref{E30}) for each quantum state,
i.e., each quantum state possesses the own interaction Hamiltonian. It is
possible to reconcile this situation with the principle of superposition and
probabilistic interpretation of wave functions analogously to the
introduction of spin into the Schr\"{o}dinger equation by Pauli \cite{R6}.

We recall that the motion of an electron in the constant homogeneous
magnetic field $H_{z}$ directed along the $z$ axis is described by two
equations with regard for the electron spin:
\begin{equation}
\left[ \hat{H}_{0}+\frac{e\hbar }{2mc}H_{z}\right] \Psi _{1}=E_{1}\Psi _{1},
\label{E55}
\end{equation}
\begin{equation}
\left[ \hat{H}_{0}-\frac{e\hbar }{2mc}H_{z}\right] \Psi _{2}=E_{2}\Psi _{2},
\label{E56}
\end{equation}
where the wave functions $\Psi _{1}$ and $\Psi _{2}$ describe the states
with the projections of the spin onto the $z$ axis $s_{z}=+\frac{1}{2}$ and $%
s_{z}=-\frac{1}{2}$, respectively. Here, $\hat{H}_{0}$ stands for the
Hamiltonian of the Schr\"{o}dinger equation for a charged particle in the
external electromagnetic field. The full wave function is two-component by
the proposition by Pauli and is written in the form of a matrix column:
\begin{equation}
{\bf \Psi }=\left(
\begin{array}{c}
\Psi _{1} \\
\Psi _{2}
\end{array}
\right) .  \label{E57}
\end{equation}
As a function ${\bf \Psi }^{\dagger }$, we choose the so-called
Hermite-conjugate wave function ${\bf \Psi }^{\dagger }=\left(
\begin{array}{cc}
\Psi _{1}^{\ast }, & \Psi _{2}^{\ast }
\end{array}
\right) $, whose elements are not only complex conjugate but also
transposed. Then the density of probability is given by
\begin{equation}
{\bf \Psi }^{\dagger }{\bf \Psi }=\left(
\begin{array}{cc}
\Psi _{1}^{\ast }, & \Psi _{2}^{\ast }
\end{array}
\right) \left(
\begin{array}{c}
\Psi _{1} \\
\Psi _{2}
\end{array}
\right) =\Psi _{1}^{\ast }\Psi _{1}+\Psi _{2}^{\ast }\Psi _{2}.  \label{E58}
\end{equation}

Consider a system with $N$ quantum states ($N$ can be infinite, and some
share of functions corresponds to a discrete spectrum whereas the rest to a
continuous one). Similarly to (\ref{E57}), we introduce wave functions, each
of them is represented by a matrix with one column and $N$ rows, where $\Psi
_{1},\ldots ,\Psi _{N}$ normed in a proper way describe possible states of
the quantum system with the energies $E_{1},\ldots ,E_{N}$, respectively:
\begin{equation}
{\bf \Psi }_{1}=\left(
\begin{array}{c}
\Psi _{1}({\bf r,R,}t) \\
0 \\
\vdots  \\
\vdots  \\
0
\end{array}
\right) ,\text{\ }{\bf \Psi }_{2}=\left(
\begin{array}{c}
0 \\
\Psi _{2}({\bf r,R,}t) \\
0 \\
\vdots  \\
0
\end{array}
\right) ,\,\ldots \,,\,{\bf \Psi }_{N}=\left(
\begin{array}{c}
0 \\
\vdots  \\
\vdots  \\
0 \\
\Psi _{N}({\bf r,R,}t)
\end{array}
\right) \ \ \ .\   \label{E59}
\end{equation}

In correspondence with such a construction of state vectors, the quantity $%
{\bf \Psi }_{k}$ is related to that situation where the system has the
energy $E_{k}$ with probability one. Then any quantum state of the
two-particle system can be presented as
\begin{equation}
{\bf \Psi }=\left(
\begin{array}{c}
a_{1}\Psi _{1}({\bf r,R,}t) \\
a_{2}\Psi _{2}({\bf r,R,}t) \\
\vdots  \\
\vdots  \\
a_{N}\Psi _{N}({\bf r,R,}t)
\end{array}
\right) ,  \label{E60}
\end{equation}
where $a_{k}$ are arbitrary numbers, complex ones in the general case. The
requirement that the state vector (\ref{E57}) be unit, i.e., the scalar
product of ${\bf \Psi }$ and the corresponding vector ${\bf \Psi }^{\dagger }
$ conjugated by Hermite be equal to 1, allows one to interpret $\left|
a_{k}\right| ^{2}$ as the probability for the system to be in the state $%
\Psi _{k}$:
\begin{equation}
\int {\bf \Psi }^{\dagger }{\bf \Psi }d\tau =\sum_{k=1}^{N}\left|
a_{k}\right| ^{2}=1.  \label{E61}
\end{equation}
Now the wave equation for the fully nonrelativistic two-body problem has the
following form:
\begin{equation}
i\hbar \frac{\partial }{\partial t}\left(
\begin{array}{c}
\Psi _{1}({\bf r,R,}t) \\
\Psi _{2}({\bf r,R,}t) \\
\vdots  \\
\vdots  \\
\Psi _{N}({\bf r,R,}t)
\end{array}
\right) =\left(
\begin{array}{ccccc}
\hat{H}(\varepsilon _{1}) & 0 & 0 & \cdots  & 0 \\
0 & \hat{H}(\varepsilon _{2}) & 0 & \cdots  & 0 \\
0 & 0 & \ddots  & 0 & \vdots  \\
\vdots  & \vdots  & 0 & \ddots  & 0 \\
0 & 0 & \cdots  & 0 & \hat{H}(\varepsilon _{N})
\end{array}
\right) \left(
\begin{array}{c}
\Psi _{1}({\bf r,R,}t) \\
\Psi _{2}({\bf r,R,}t) \\
\vdots  \\
\vdots  \\
\Psi _{N}({\bf r,R,}t)
\end{array}
\right) ,  \label{E62}
\end{equation}
where $\Psi _{k}$ is the normed wave function of the $k$ quantum state and
\begin{equation}
\hat{H}(\varepsilon _{k})={\left[ -\frac{\hbar ^{2}}{2M}\Delta _{{\bf R}}-%
\frac{\hbar ^{2}(1-\varepsilon _{k})^{2}}{2\mu }\Delta _{{\bf r}}+V(r)\right]
,}  \label{E63}
\end{equation}
\begin{equation}
\varepsilon _{k}=\left\langle \Psi _{k}\left| \hat{\varepsilon}(\left| {\bf F%
}_{12}(r)\right| )\right| \Psi _{k}\right\rangle \,.  \label{E64}
\end{equation}
We recall that the average value of any function $g$ of the operators ${\bf
\hat{r}}_{1}$, ${\bf \hat{r}}_{2}$, ${\bf \hat{p}}_{1}$, and ${\bf \hat{p}}%
_{2}$ has the following form:
\begin{equation}
\int {\bf \Psi }^{\dagger }\text{ }{\bf \hat{g}}\text{ }{\bf \Psi }d\tau
=\sum_{k=1}^{N}\left| a_{k}\right| ^{2}\int \Psi _{k}^{\ast }\text{ }g[{\bf
\hat{r}}_{1},{\bf \hat{r}}_{2},{\bf \hat{p}}_{1}(\varepsilon _{k}),{\bf \hat{%
p}}_{2}(\varepsilon _{k})]\text{ }\Psi _{k}d\tau ,  \label{E65}
\end{equation}
where the matrix ${\bf \hat{g}}$ is diagonal
\begin{equation}
{\bf \hat{g}}=\left(
\begin{array}{ccccc}
g(\varepsilon _{1}) & 0 & 0 & \cdots  & 0 \\
0 & g(\varepsilon _{2}) & 0 & \cdots  & 0 \\
0 & 0 & \ddots  & 0 & \vdots  \\
\vdots  & \vdots  & 0 & \ddots  & 0 \\
0 & 0 & \cdots  & 0 & g(\varepsilon _{N})
\end{array}
\right) ,  \label{E66}
\end{equation}
and $g(\varepsilon _{k})\equiv g[{\bf \hat{r}}_{1},{\bf \hat{r}}_{2},{\bf
\hat{p}}_{1}(\varepsilon _{k}),{\bf \hat{p}}_{2}(\varepsilon _{k})]$. In (%
\ref{E65}), the summation is carried on over all discrete states of the
system. If a continuous spectrum is available, we perform the summation over
the entire discrete spectrum and the corresponding integration over the
whole continuous spectrum in (\ref{E65}).

If all $\varepsilon _{k}$ tend to zero, all $\hat{H}(\varepsilon _{k})$
become identical and equal to the Hamiltonian of the Schr\"{o}dinger
equation, and all wave functions are the eigenfunctions of this Hamiltonian.

\section{Discrete spectrum of a hydrogenlike atom within the model of
noncommuting operators of coordinates and momenta of different particles}

Consider the discrete spectrum of a hydrogenlike atom. Let two particles
with masses $m_{1}$ (electron) and $m_{2}$ (atomic nucleus) be bound by the
Coulomb potential $V(r)=-Ze^{2}r^{-1}$, where $Z$ is the charge of the
atomic nucleus. The self-consistent system of integro-differential equations
for every state with the binding energy $E_{nl}$ (\ref{E38})-(\ref{E39}) can
be written as
\begin{equation}
\left[ -\frac{\hbar ^{2}(1-\varepsilon _{nl})^{2}}{2\mu }\left( \frac{d^{2}}{%
dr^{2}}-\frac{l(l+1)}{r^{2}}\right) -\frac{Ze^{2}}{r}\right] \chi
_{nl}(r)=E_{nl}\chi _{nl}(r){\bf ,}  \label{E67}
\end{equation}
\begin{equation}
\varepsilon _{nl}=\int_{0}^{\infty }\chi _{nl}^{2}(r)\frac{{\displaystyle{%
\frac{Ze^{2}}{r^{2}}}}}{{\displaystyle{\frac{Ze^{2}}{r^{2}}}}+{\displaystyle{%
\frac{\left( \mu c^{2}\right) ^{2}}{\hbar c}}}}dr,  \label{E68}
\end{equation}
\begin{equation}
\int_{0}^{\infty }\chi _{nl}^{2}(r)\,dr=1.  \label{E69}
\end{equation}

Two equations (\ref{E67}) and (\ref{E69}) are the equations for the normed
radial functions of a hydrogenlike atom by the Schr\"{o}dinger theory. Their
solutions for bound states are well-known (see, e.g., \cite{R7}):
\begin{equation}
\chi _{nl}(r)=N_{nl}r^{l+1}F\left( -n+l+1,2l+2,\frac{2Zr}{(1-\varepsilon
_{nl})^{2}na_{0}}\right) \exp \left( \frac{-Zr}{(1-\varepsilon
_{nl})^{2}na_{0}}\right) \,,  \label{E70}
\end{equation}
where
\begin{equation}
N_{nl}=\frac{1}{(2l+1)!}\left[ \frac{(n+l)!}{2n(n-l-1)!}\right] ^{1/2}\left(
\frac{2Z}{(1-\varepsilon _{nl})^{2}na_{0}}\right) ^{l+3/2}\,.  \label{E71}
\end{equation}
Here, $a_{_{0}}=\hbar ^{2}/\mu e^{2}$ is the Bohr radius, and $F$ is a
degenerate hypergeometric function. Eigenenergies are
\begin{equation}
E_{nl}=-\frac{\mu c^{2}}{2}\frac{(\alpha Z)^{2}}{n^{2}}\frac{1}{%
(1-\varepsilon _{nl})^{2}}\,.  \label{E72}
\end{equation}
Here, $\alpha =e^{2}/\hbar c$ is the fine structure constant, $l=0,1,\ldots
,n-1,$ and $n=1,2,\ldots ,\infty $. By substituting $\chi _{nl}(r)$ into
Eq.~(\ref{E68}), we obtain the nonlinear equation for the determination of $%
\varepsilon _{nl}:$%
\begin{equation}
\eta _{nl}=S_{nl}\int_{0}^{\infty }x^{2l+2}\exp \left( -x\right) F^{2}\left(
-n+l+1,2l+2,x\right) \left( 1+\frac{4(\alpha Z)^{3}}{n^{2}\eta _{nl}^{4}x^{2}%
}\right) ^{-1}\,dx\,,  \label{E73}
\end{equation}
where $\eta _{nl}=1-\varepsilon _{nl}$ and $S_{nl}=\left[ (2l+1)!\right]
^{-2}{[2n(n-l-1)!]}^{-1}\left[ (n+l)!\right] .$

For a hydrogenlike atom in the ground state, we have:
\begin{equation}
\eta _{10}=\frac{1}{2}\int_{0}^{\infty }x^{2}\exp (-x)\,\left( 1+\frac{%
4(\alpha Z)^{3}}{\eta _{10}^{4}x^{2}}\right) ^{-1}dx\,.  \label{E74}
\end{equation}
With respect to $\eta _{10}$, this equation has solutions if $\alpha Z\leq
\alpha Z_{c}=0.510107$ (Fig. \ref{fig1}). From two solutions, a solution
being closer to unit is suitable. The second should be omitted since it
corresponds to the case where the binding energy tends to minus infinity as
the interaction constant $\alpha Z$ tends to zero, which is physically
unacceptable. For $Z>Z_{c}$, Eq. (\ref{E74}) has no solutions, which means
that the given bound state does not exist.

Fig. \ref{fig2} displays the binding energy of the ground state of a
hydrogenlike atom along with the analogous dependence by the Schr\"{o}dinger
theory. For the sake of comparison, we also present the corresponding
solution of the relativistic Klein-Gordon equation for a spinless particle
with mass $\mu $ in the Coulomb field $V(r)=-Ze^{2}r^{-1}$ as
\begin{equation}
E_{nl}=\mu c^{2}\left\{ -1+\left[ 1+\alpha ^{2}Z^{2}\left( n-l-0.5+\left[
\left( l+0.5\right) ^{2}-\alpha ^{2}Z^{2}\right] ^{1/2}\right) ^{-2}\right]
^{-1/2}\right\} \,.  \label{E75}
\end{equation}

We see that the energy levels are situated below the Schr\"{o}dinger ones
and higher than that computed according to the Klein-Gordon theory.
Analogous calculations can be easily performed for excited states of a
hydrogenlike atom (Figs. \ref{fig3},\ref{fig4}). In this case, it turns out
that the Schr\"{o}dinger levels with given $n$ are split into $n$ closely
positioned sublevels since the orbital number $l$ can take $n$ values ($%
l=0,1,\ldots ,n-1$), i.e., the degeneracy is removed. All levels with a
given $n$ and different $l$ are situated below the corresponding
Schr\"{o}dinger level.

For a hydrogen atom, the parameter of noncommutativity $\varepsilon _{nl}$
is significantly less than 1 ($\varepsilon _{10}=0.776\times
10^{-6},\varepsilon _{20}=0.970\times 10^{-7},\varepsilon _{21}=0.324\times
10^{-7}$). The splitting of the level with $n=2$ is also very small and is
by two orders of magnitude less than that by the Dirac theory for a hydrogen
atom.

The parameter of noncommutativity $\varepsilon $ of the operators of
coordinates and momenta of different particles , presented in Fig. \ref{fig5}%
, decreases as the quantum numbers $n$ and $l$ increase (for the same $Z$),
i.e., fully nonrelativistic solutions pass into solutions of the
Schr\"{o}dinger equation for large quantum numbers.

A characteristic feature of the fully nonrelativistic equation is the
presence of the critical value of the parameter $\alpha Z_{c}$ for each
energy level, which is lacking in the case of the nonrelativistic
Schr\"{o}dinger equation. For example, for levels with $n=2$, $\alpha
Z_{c}=1.401098$ for $l=0$ and $\alpha Z_{c}=1.221611$ for $l=1$.

As the parameter $\alpha Z$ grows, the average distance between particles
decreases. For the ground state of a hydrogenlike atom, $\langle |{\bf \hat{r%
}}_{2}-{\bf \hat{r}}_{1}|\rangle =3\hbar (1-\varepsilon _{10})^{2}(2\mu
c\alpha Z)^{-1}$ \ and takes the minimum value $\langle |{\bf \hat{r}}_{2}-%
{\bf \hat{r}}_{1}|\rangle \approx 1.33\hbar /\mu c$ for $Z=Z_{c}$. With a
further increase in the parameter $\alpha Z$, the self-consistent system of
equations (\ref{E67})-(\ref{E69}) has no solutions in the state with $n=1$,
i.e., the 1S-state cannot exist, and the ground state is the state with $n=2$
and $l=0$ (2S-state) if $0.510<\alpha Z<0.847$ or $1.222<\alpha Z\leq 1.401$%
. In the region $0.847<\alpha Z\leq 1.222$, the ground state is the state
with $n=2$ and $l=1$ since $\varepsilon _{21}>\varepsilon _{20}$ (2P-state).
Here, we observe the substantial distinction from solutions of the
nonrelativistic Schr\"{o}dinger equation, for which the ground state is, as
known, the 1S-state. Solutions of the proposed fully nonrelativistic
solution for a hydrogenlike atom are also significantly different from
solutions of the Klein-Gordon equation. The latter has the critical value of
the interaction constant $\alpha Z_{c}=0.5$. Above it, the continuous energy
spectrum appears, and there ''occurs'' the so-called fall to the center.

Now we write down the values of quantum Poisson brackets proposed by Dirac
\cite{R8}:
\begin{equation}
\{\hat{x}_{1},\,\hat{p}_{1x}\}=1-\frac{m_{2}}{M}\varepsilon _{nl}\,,
\label{E76}
\end{equation}
\begin{equation}
\{\hat{x}_{2},\,\hat{p}_{2x}\}=1-\frac{m_{1}}{M}\varepsilon _{nl}\,,
\label{E77}
\end{equation}
\begin{equation}
\{\hat{x}_{1},\,\hat{p}_{2x}\}=\frac{m_{2}}{M}\varepsilon _{nl}\,,
\label{E78}
\end{equation}
\begin{equation}
\{\hat{x}_{2},\,\hat{p}_{1x}\}=\frac{m_{1}}{M}\varepsilon _{nl}\,,
\label{E79}
\end{equation}
\begin{equation}
\{\hat{x}_{1},\,\hat{x}_{2}\}=0\,,  \label{E80}
\end{equation}
\begin{equation}
\{\hat{p}_{1x},\,\hat{p}_{2x}\}=0\,.  \label{E81}
\end{equation}
As $\varepsilon _{nl}\,\rightarrow 0$, these brackets are transformed into
the classical Poisson brackets. That is, in this case, we have a complete
analogy between classical mechanics and quantum mechanics. Fig. \ref{fig5}
demonstrates that $\varepsilon $ significantly differs from zero for
systems, whose size is about the Compton wavelengths of particles composing
the system. In this case, there is no similar analogy with classical
mechanics.

\section{Conclusion}

The Schr\"{o}dinger equation for a system of interacting particles is not a
strictly nonrelativistic equation because it is grounded on the implicit
assumption about finiteness of the interaction propagation velocity. The
last means that if the commutator of operators of a coordinate and the
corresponding momentum of a free particle is defined as
\begin{equation}
\left[ \hat{x},\,\hat{p}_{x}\right] =i\hbar \,,  \label{E82}
\end{equation}
this commutator for a system of coupled particles has the same value $i\hbar
\,$.$\,$However, in a nonrelativistic quantum system during measurement of
the coordinate of a particle, a whole transferred momentum is distributed
over all particles but is not transferred to only the measured one.
Therefore, in a system of interacting particles, this commutator should have
the form
\begin{equation}
\left[ \hat{x},\,\hat{p}_{x}\right] =i\hbar \delta \,,  \label{E83}
\end{equation}
where $0<\delta \leq 1$ .

The rejection of the implicit assumption on finiteness of the propagation
velocity of interactions implies the noncommutativity of the operators of
coordinates and momenta of different particles. But the operators of
coordinates of all particles and operators of momenta of all particles
mutually commute that allows one to use these collections as independent
variables.

The derived self-consistent system of integro-differential equations allows
one to separate the motion of the center of mass of the system, which moves
as a free particle.

The properties of solutions of the proposed system of equations
significantly differ from those of Schr\"{o}dinger solutions for systems,
whose size is comparable with the Compton wavelength of particles. That is,
the consideration of noncommutativity of the operators of coordinates and
momenta of different particles is important for the quantum mechanics of
atoms with a large charge of nuclei as well as for the phenomena of nuclear
physics, for which the size of a system is about the Compton wavelength of
particles composing the system.

In conclusion, the author expresses his gratitude to Dr. V. V. Kukhtin and
Prof. I. V. Simenog for a very useful discussion of certain problems touched
in this work.

\begin{figure}[tbp]
\caption{Dependence of the right-hand side of (74) on $\protect\eta $ for
various values of the parameter $\protect\alpha Z$.}
\label{fig1}
\end{figure}
\begin{figure}[tbp]
\caption{Binding energy $E_{10}$ of the ground state of a hydrogenlike atom
(72) vs the parameter $\protect\alpha Z$. The upper dotted line corresponds
to the Schr\"{o}dinger theory, $E_{S}=-\protect\mu c^{2}(\protect\alpha
Z)^{2}/2$, and the lower dotted line to the corresponding solution of the
Klein-Gordon equation (75).}
\label{fig2}
\end{figure}
\begin{figure}[tbp]
\caption{Binding energy $E_{20}$ of a hydrogenlike atom (72) vs the
parameter $\protect\alpha Z$. The upper dotted line corresponds to the
Schr\"{o}dinger theory, $E_{S}=-\protect\mu c^{2}(\protect\alpha Z)^{2}/8$,
and the lower dotted line to the corresponding solution of the Klein-Gordon
equation (75).}
\label{fig3}
\end{figure}
\begin{figure}[tbp]
\caption{Binding energy $E_{21}$ of a hydrogenlike atom (72) vs the
parameter $\protect\alpha Z$. The upper dotted line corresponds to the
Schr\"{o}dinger theory, $E_{S}=-\protect\mu c^{2}(\protect\alpha Z)^{2}{/8}$%
, and the lower dotted line to the corresponding solution of the
Klein-Gordon equation (75).}
\label{fig4}
\end{figure}
\begin{figure}[tbp]
\caption{ Dependence of the parameter of noncommutativity of operators $%
\protect\varepsilon $ for the lowest states of a hydrogenlike atom vs the
interaction constant $\protect\alpha Z$.}
\label{fig5}
\end{figure}

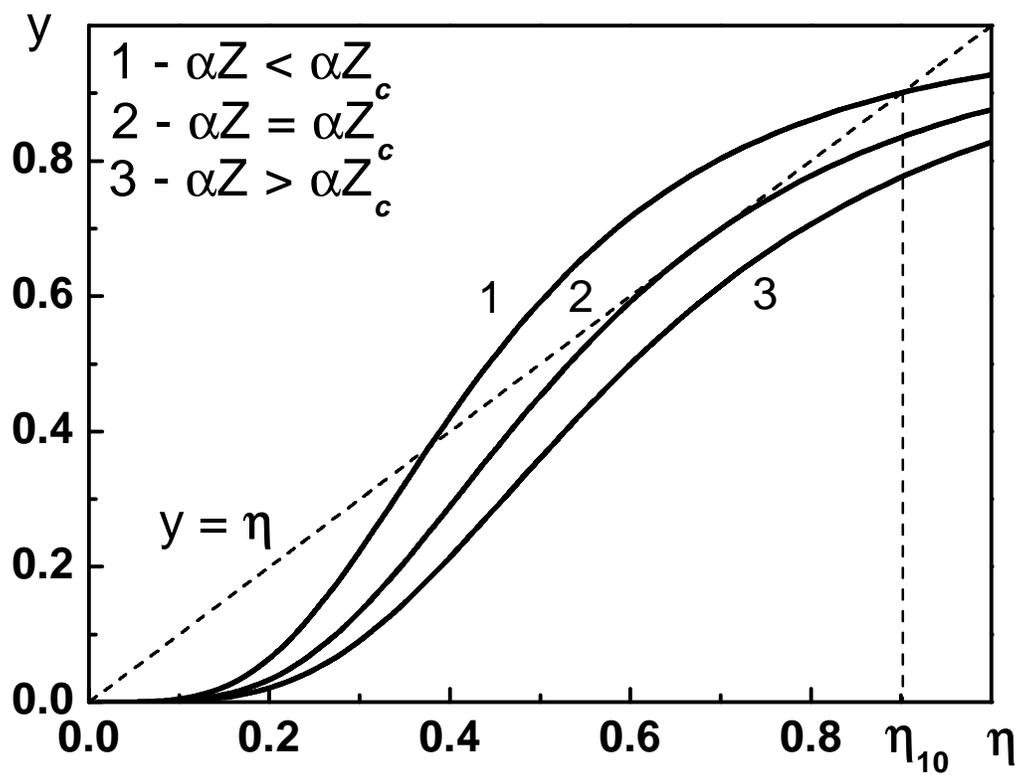
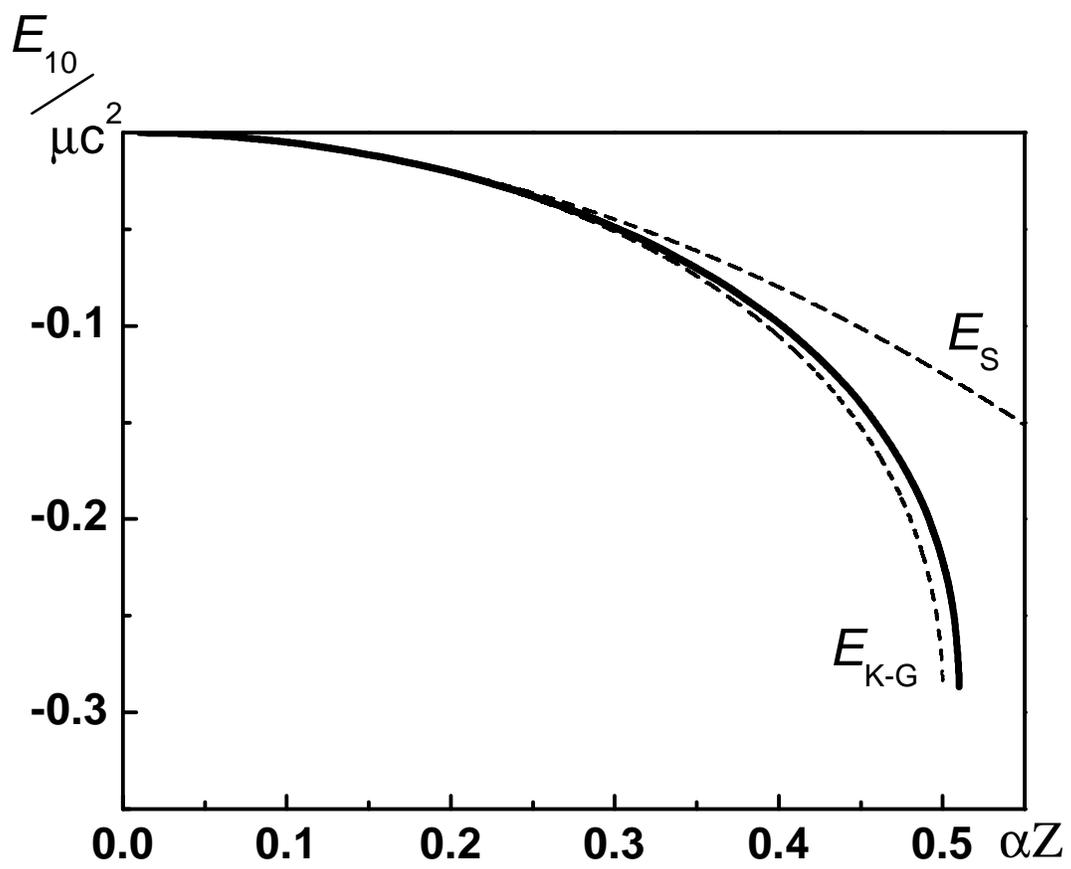
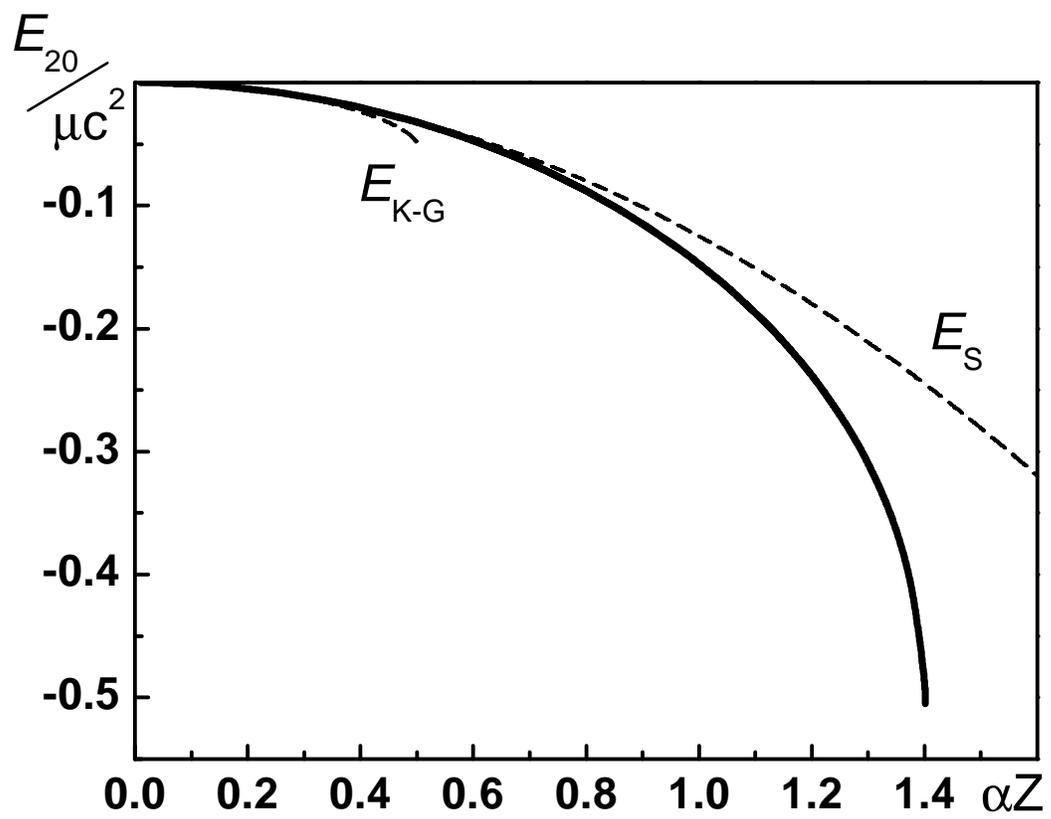
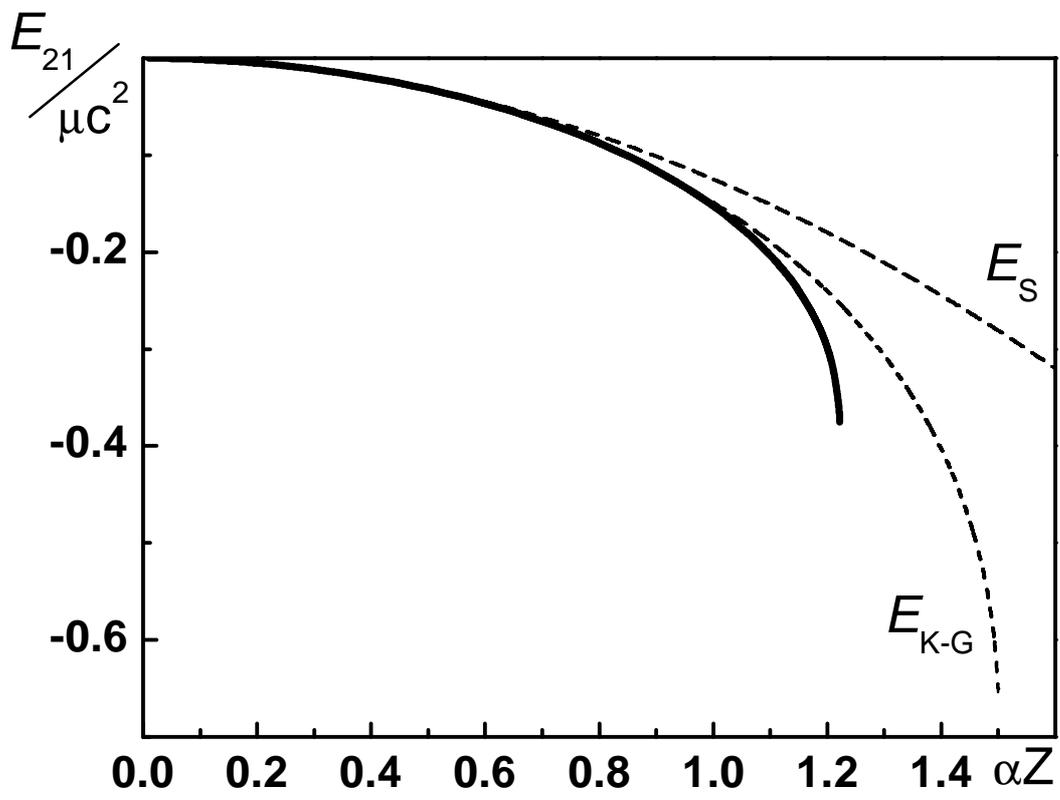
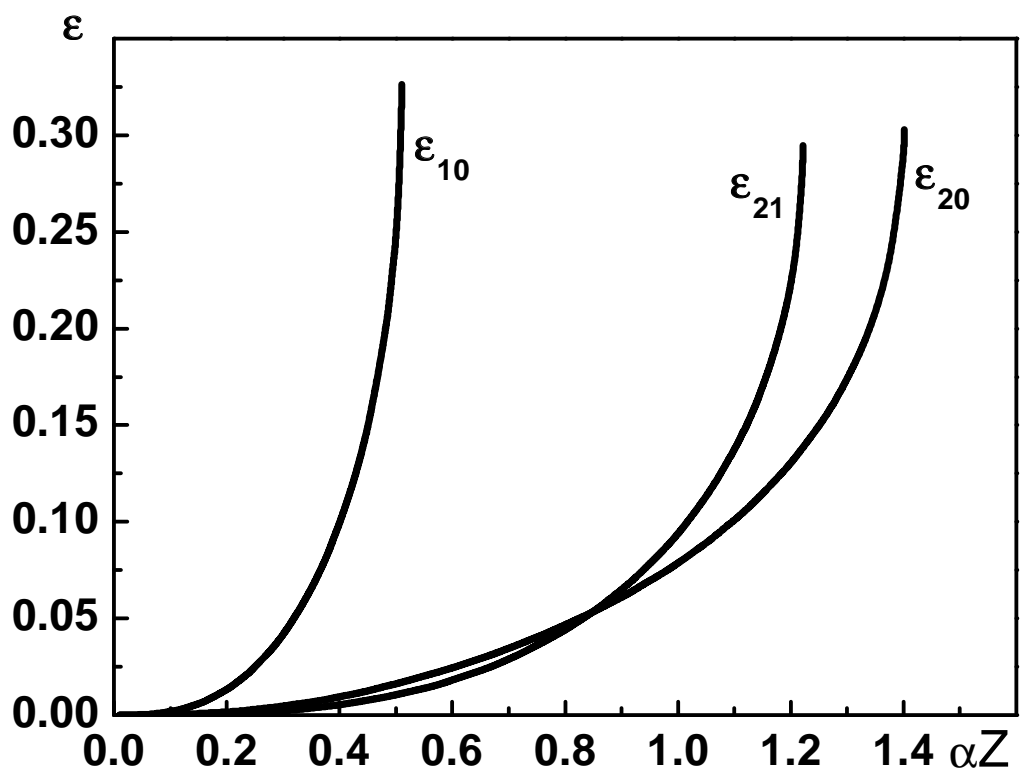
\end{document}